\documentclass[runningheads]{llncs}
\usepackage[T1]{fontenc}
\usepackage{graphicx}

\input{macro}
\begin{document}
\title{Verified and Optimized Implementation of Orthologic Proof Search}

\author{Simon Guilloud\orcidID{0000-0001-8179-7549} and Clément Pit-Claudel\orcidID{0000-0002-1900-3901}}
\institute{EPFL \\ School of Computer and Communication Sciences \\ Station 14, CH-1015 Lausanne, Switzerland \\ 
\email{\{firstname.lastname\}@epfl.ch}}

\authorrunning{S. Guilloud and C. Pit-Claudel}

\sloppy

\maketitle

\begin{abstract}
    We report on the development of an optimized and verified decision procedure for orthologic equalities and inequalities. This decision procedure is quadratic-time and is used as a sound, efficient and predictable approximation to classical propositional logic in automated reasoning tools. We formalize, in the Coq proof assistant, a proof system in sequent-calculus style for orthologic. We then prove its soundness and completeness with respect to the algebraic variety of ortholattices, and we formalize a cut-elimination theorem (in doing so, we discover and fix a missing case in a previously published proof).  
    
    We then implement and verify a complete proof search procedure for orthologic. A naive implementation is exponential, and to obtain an optimal quadratic runtime, we optimize the implementation by memoizing its results and simulating reference equality testing. We leverage the resulting correctness theorem to implement a reflective Coq tactic. We present benchmarks showing the procedure, under various optimizations, matches its theoretical complexity. 
    
    Finally, we develop tactics including normalization with respect to orthologic and a boolean solver, which we also benchmark. We make tactics available as a standalone Coq plugin.
\end{abstract}

\section{Introduction}
Specialized, reliable and efficient building blocks are indispensable in scaling automated reasoning software. Program verifiers, SMT solvers, proof assistants and automated theorem provers use them to tackle the various theories and subproblems that compounds a logical statement. As they combine, so do the possibility of an implementation error. To ensure that they can be used as trusted components in program verification pipelines, it is very desirable to verify these verification algorithms themselves.

One fragment of particular interest is propositional logic. Despite significant progress in SAT solvers, solving satisfiability or validity of propositional formulas remains a major challenge to scalability of decision procedures. An alternative and complementary approach to heuristics is orthologic-based reasoning. Orthologic is a non-distributive generalization of classical propositional logic which admits polynomial-time ($\mathcal{O}(n^2)$) validity checking and normalization algorithms \cite{brunsFreeOrtholattices1976,guilloudFormulaNormalizationsVerification2023,guilloudOrthologicAxioms2024}. Orthologic offers a trade-off: it sacrifices completeness (with respect to classical semantics) in exchange for guaranteed efficiency and predictability. It also admits a quadratic-time normalization algorithm, allowing to compute a unique normal form (with respect to the laws of orthologic) of guaranteed minimal size. This helps with caching, solves a variety of intuitive formula equivalences, and allows simplifying formulas before solving them with more complete but possibly inefficient procedures. Recent work has demonstrated the practical utility of orthologic-based reasoning, being an important component of verification tools such as the Stainless program verifier  \cite{guilloudFormulaNormalizationsVerification2023} and the Lisa proof assistant \cite{guilloudLISAModernProof2023}. \autoref{tab:algebraiclaws} presents the laws of orthologic.

\begin{table*}[bth]    
    \centering
    \begin{tabular}{r c @{\hskip 2em} | @{\hskip 2em} r c}
         V1: & $x \lor y = y \lor x$  & V1': & $x \land y = y \land x$ \\
         V2: & $x \lor ( y \lor z) = (x \lor y) \lor z$  & V2': & $x \land ( y \land z) = (x \land y) \land z$ \\
         V3: & $x \lor x = x$  & V3': & $x \land x = x$ \\
         V4: & $x \lor 1 = 1$  & V4': & $x \land 0 = 0$ \\
         V5: & $x \lor 0 = x$  & V5': & $x \land 1 = x$ \\
         V6: & $\neg \neg x = x$\\
         V7: & $x \lor \neg x = 1$  & V7': & $x \land \neg x = 0$ \\
         V8: & $\neg (x \lor y) = \neg x \land \neg y$  & V8': &  $\neg (x \land y) = \neg x \lor \neg y$ \\
         V9: & $x \lor (x \land y) = x$ & V9': & $x \land (x \lor y) = x$   \\
    \end{tabular}
    
    \
    \caption{Laws of ortholattices, algebraic varieties with signature $(S, \land, \lor, 0, 1, \neg)$.}
    \label{tab:algebraiclaws}
\end{table*}

In this work, we present the first formalization and verification, using the Coq proof assistant, of an efficient decision procedure for the validity and equivalence problems for propositional formulas in orthologic. This algorithm \cite{guilloudOrthologicAxioms2024} is based on proof search in a proof system for orthologic that is a restriction of the classical sequent calculus where a sequent can never contain more than two distinct formulas at a time. Despite this restriction, the proof system admits cut elimination, which we formalize. The cut elimination property implies a subformula property, which is key to the completeness of the proof search procedure. 

A naive implementation of the decision procedure has exponential complexity. To obtain a polynomial version, we leverage memoization, i.e. storing in a table the intermediate results of the recursive calls. Moreover, using structural equality to check if a key is in the memoization map costs an additional linear runtime factor. 
To obtain an optimal quadratic version, we modify the algorithm to use a form of reference equality. 
As our algorithm is purely functional, this means we have to extend our formulas abstract syntax trees to assign to each node a unique identifier (or pointer) that can then be used in the memoization map. We formally prove the correctness of these constructions. 

We implement and verify multiple versions of the algorithm: without optimization ($\widetilde{\mathcal O} (2^n)$), with memoization using lists ($\widetilde{ \mathcal O} (n^5)$), using AVL maps ($\widetilde{ \mathcal O} (n^3)$), and using AVL maps and simulated reference equality ($\widetilde{ \mathcal O} (n^2)$).
Memoization and reference equality are generic, and important in many tools and algorithms. Our approach to verifying these optimizations is not specific to orthologic proof search and we expect it to extend to any recursive algorithms over algebraic data types.

Using the technique of proofs by reflection, we then obtain a set of executable proof tactics that decides equality modulo orthologic rule and which is applicable to any ortholattice, in particular the type \lstinline|bool| of boolean values. This tactic is able to solve automatically, for example,
\begin{lstlisting}
    true = (a && b) || (negb a) || (negb b)
\end{lstlisting}
. We present benchmarks attesting that each version of the proof search procedure (with memoization using a list-based map, with memoization using AVL trees, and with both memoization and reference equality) meets the theoretical complexity.

Many formulas are not completely provable in the fragment of orthologic, in which case the decision procedure alone cannot do anything, even if most of the structure of the formula reduces under orthologic rules. The orthologic simplification algorithm \cite{guilloudFormulaNormalizationsVerification2023} on the other hand is widely applicable, and can make substantial progress on a problem even if the formula is not equivalent to true. This is also very useful if the propositional formulas are built on some theories, such as arithmetic, in which case even classical boolean decision procedures would not work. We implement this simplifier in OCaml as a Coq plugin, using the above tactic to efficiently prove equivalence with the original expression. This yields a generic simplification tactic for terms of type \lstinline|bool|.

We finally leverage this normalization tactic to implement a boolean solver which alternates between branching on a variable and normalizing the formulas, as in \cite{guilloudLISAModernProof2023}. We present benchmarks comparing this procedure to the built-in \lstinline|btauto| tactic.

All our tactics are available as a Coq plugin alongside with benchmarks at [redacted].

\subsection{Contributions}

In \autoref{sect:orthologic}, we first formalize in the Coq proof assistant a proof system for orthologic (in the style of sequent calculus) and its soundness and completeness with respects to truth in every ortholattice.
In \autoref{sect:cutelim}, we then formalize the cut elimination theorem (without non-logical axioms) of orthologic, and in the process discover a missing edge case in the proof of \cite{guilloudOrthologicAxioms2024}.

In \autoref{sect:tactic}, we implement the proof-search based decision procedure for orthologic from \cite{guilloudOrthologicAxioms2024}. We prove soundness and completeness of the algorithm. We then use reflection to obtain a Coq proof tactic which can solve any equality or inequality valid in orthologic, for any ortholattice, in particular the type \lstinline|bool| of truth values.
In \autoref{sect:memoization}, we describe our implementation and verification of the memoized version of the algorithm. 
In \autoref{sect:pointers}, we discuss reference equality. This optimization is significantly more complex. It requires redefining trees representing orthologic terms to include pointers and proving that the transformation from pointer-free terms to terms with pointers is correct, which was surprisingly difficult. 

In \autoref{sect:ocaml}, we present an alternative tactic that directly produces proof terms rather than using reflection, which is faster in practice but less space efficient. We also implement tactics for OL-normalization and an orthologic-based boolean solver. 

In \autoref{sect:bench}, we present a series of benchmark which are supposed to hit the asymptotic worst case of every implementation of the proof search procedure, using different reduction strategies, and find that the resulting curves match the theoretical complexity. We then present benchmark comparing our OL-based boolean validity tactic to the built-in \lstinline|btauto| tactic of Coq on random formulas generated according to \cite{navarroGenerationHardNonclausal2005}.

\subsection{Related Work}
The word problem for ortholattices (deciding if given two terms, one is always $\leq$ than the other) was first solved by \cite{brunsFreeOrtholattices1976}, extended in \cite{guilloudFormulaNormalizationsVerification2023} to obtain normal forms of terms. A similar proof system for orthologic, with the property that sequents never contain more than one formula, was already described by \cite{schultemontingCutEliminationWord1981}. Different proof systems have also been considered \cite{meinanderSolutionUniformWord2010,laurentFocusingOrthologic2016,kawanoLabeledSequentCalculus2018}.

The authors of \cite{forsterCompletenessTheoremsFirstorder2021} formalize cut elimination for intuitionistic logic and \cite{fruminSemanticCutElimination2021} cut elimination for the logic of bunched implications in Coq, but their approach is semantic rather than syntactic. 
\cite{tewsFormalizingCutElimination2013}, still using Coq, show Cut elimination for a class of coalgebraic logics, including a variety of modal logics. They propose both a syntactic and a semantic proof. In Isabelle, \cite{urbanRevisitingCutEliminationOne2008} formalizes a particular, strongly normalizing version of cut elimination for classical first order logic.

Whitman's algorithm, a decision procedure for inequalities holding in every lattice (without negation), has been formalized in Coq by \cite{jamesReflectionbasedProofTactic2009}, also using reflection.

Ortholattices have been defined in Mizar \cite{truszkowskaTwoShortAxiomatizations}, but little about them proven. In Coq, \cite{laurentFocusingOrthologic2016} formalized results about a different proof system for orthologic.

Verified memoization has been studied in \cite{wimmerVerifiedMemoizationDynamic2018}. The authors propose a framework for automatic verified memoization of programs in Isabelle/HOL. In Coq, \cite{braibantImplementingReasoningHashconsed2014} studies the related topic of hash-consing, which can be seen as memoization of constructors. In their approach, they entirely replace nodes of an ADT by identifiers, so that the recursive structure only exist in the  hash-consing map, while we add the identifier to the tree.

\section{Formalizing Ortholattices and Orthologic}
\label{sect:orthologic}

We first formalize the algebraic class of ortholattices as a typeclass (\autoref{lst:ortholattice}).
\begin{figure*}[t]
\begin{lstlisting}[label=lst:ortholattice,caption=Definition of an Ortholattice]
Class Ortholattice := {
  A : Set;
  leq : relation A where "x <= y" := (leq x y);
  meet : A -> A -> A where "x ∩ y" := (meet x y);
  join : A -> A -> A where "x ∪ y" := (join x y);
  neg : A -> A where "¬ x" := (neg x);
  zero : A; one : A;
  ...
}.
\end{lstlisting}
\end{figure*}
Ortholattices are of course \textit{lattices}, and in particular a partial order with $\leq$ given by 
$$x \leq y \iff x \land y = x$$
or equivalently
$$x \leq y \iff x \lor y = y.$$
Ortholattices can be axiomatized equivalently as an algebraic variety, such as in \autoref{tab:algebraiclaws}, or as a partial order.

We then implement with Ltac \textit{Whitman's algorithm} \cite{freeseFreeLattices1995}, a simple decision procedure for lattices. This helps us to quickly show a number of useful lemmas about ortholattices.
\iffalse
\begin{lstlisting}
Ltac Whitman := match goal with
  | [ |- _ == _] => rewrite equiv_leq; split; Whitman
  | [ |- ?x <= ?x] => apply P1
  | [ |- _ <= _ ∩ _] => apply P6; Whitman
  | [ |- _ ∪ _ <= _] => apply P6'; Whitman
  | [ |- _ ∩ _ <= _] => 
    try (apply glb1; Whitman; eauto; fail);
    try (apply glb2; Whitman; eauto; fail)
  | [ |- _ <= _ ∪ _] => 
    try (apply lub1; Whitman; eauto; fail);
    try (apply lub2; Whitman; eauto; fail)
  | [ |- ¬ _ <= ¬ _] => apply P8; Whitman
  | [ |- _ <= _] => try (eauto;fail)
end.
Lemma example {OL: Ortholattice} a b : (a ∩ b) <= (a ∪ b).
Proof. Whitman. Qed.
\end{lstlisting}
\fi
This simple tactic is subsumed by the tactic for ortholattices we will obtain with reflection. 
We then show that \lstinline{equiv} is a congruence relation for \lstinline|<=|, \lstinline|∩|, \lstinline|∪| and \lstinline|¬|. This makes ortholattices \textit{setoids}, and enables the use of \textit{generalized rewriting} \cite{sozeauNewLookGeneralized2009}.

\subsection*{Orthologic}
We define in the standard way the type of ortholattice terms, and the evaluation of a term in an arbitrary ortholattice:

\begin{lstlisting}
Inductive Term : Set :=
  | Var : positive -> Term
  | Meet : Term -> Term -> Term
  | Join : Term -> Term -> Term
  | Not : Term -> Term.

Fixpoint eval {OL: Ortholattice} 
  (t: Term) (f: nat -> A) : A := ...
\end{lstlisting}
%%%%%%%%%%%%%

and then implement the OL proof system of \cite{guilloudOrthologicAxioms2024}, formulated as a sequent calculus. One can think of this proof system as Gentzen's sequent calculus for classical logic~\cite{gentzenUntersuchungenUberLogische1935} restricted to ensure that at any given point in a proof, a sequent never has more than two formulas on both sides combined. 

We represent sequents as (ordered) pairs of annotated formulas as in \cite{guilloudOrthologicAxioms2024}:
\begin{lstlisting}
Inductive AnTerm : Set :=
  | N : AnTerm
  | L : Term -> AnTerm
  | R : Term -> AnTerm.
Definition Sequent (l r : AnTerm) := (l, r).
\end{lstlisting}
Where \lstinline|N| represents no formula, \lstinline|L| a formula on the left and \lstinline|R| a formula on the right.
For example, $(L\, \phi, R\, \psi)$ stands for $\phi \vdash \psi$ in more conventional notation.

We implemented the proof system using dependent inductive types, so that the correctness of a proof is guaranteed by construction, and no additional proof-checking function is required.
For example, the \texttt{RightAnd} rule from \cite{guilloudOrthologicAxioms2024}, is
\begin{center}
    \AxiomC{$\Gamma, \phi^R$}
    \AxiomC{$\Gamma, \psi^R$}
    \RightLabel{\text{ RightAnd}}
    \BinaryInfC{$\Gamma, (\phi \land \psi)^R$}
    \DisplayProof
\end{center}
and is encoded as
\begin{lstlisting}
Inductive OLProof : AnTerm*AnTerm -> Set := 
  ...
  | RightAnd: forall {a} {b} {g}, 
      OLProof (g, R a) -> OLProof (g, R b) -> OLProof (g, R (Meet a b))
  ...
\end{lstlisting}
In \cite{guilloudOrthologicAxioms2024}, sequents are formally considered as sets. We have defined them in Coq using ordered pairs and hence needs to define two additional rules, simulating the set-like nature of sequents: the Swap and Contract rules.
 \begin{center}
\begin{tabular}{l l}
    \AxiomC{$\Gamma, \Delta$}
    \RightLabel{\text{ Swap}}
    \UnaryInfC{$\Delta, \Gamma$}
    \DisplayProof
    \hspace{1em}
     & 
    \hspace{1em}
    \AxiomC{$\Gamma, \Gamma$}
    \RightLabel{\text{ Contract}}
    \UnaryInfC{$\Gamma, N$}
    \DisplayProof
\end{tabular}
\end{center}

Soundness of the proof systems states that if a sequent is provable, then the corresponding inequality must be true in every ortholattice (semantic truth), and completeness is the converse. Both are straightforward to prove.

\section{Cut Elimination for Orthologic}
\label{sect:cutelim}
The key property of the orthologic proof system is that it admits cut elimination: any provable sequent can be proved without using the cut rule. Cut elimination has important theoretical and practical consequences. Since the \lstinline|Cut| step is the only one in which a term can appear in the premise but not the conclusion, cut elimination implies the subformula property: if a sequent has a proof, then it has a proof where only subterms of the conclusion appear. This is key to the orthologic proof search procedure.

\begin{lstlisting}
Theorem cut_elimination  s (proof : OLProof s):
    {p: OLProof s | is_cut_free  p}.
\end{lstlisting}

This theorem is not straightforward to formalize. The paper proof starts with ``Consider the topmost instance of the cut rule in the proof''. This deceptively short step corresponds to doing a double induction, first on the number of cut appearing in a proof (\lstinline|fuelCut|), then on the size of the proof (\lstinline|fuelSize|)\footnote{Two nested inductions on natural numbers is equivalent to transfinite induction on $\omega^2$}.  The proof takes as arguments proofs that the given fuel is larger than the metric it represents, and every induction step need to justify that the measures are decreasing.
The problem then reduces to eliminating a single Cut step from an otherwise cut free proof.

\newcommand{\A}{\mathcal A}
\newcommand{\B}{\mathcal B}

Graphically, we have the following situation:
\begin{center}
    \AxiomC{$\A$}
    \UnaryInfC{$\Gamma, b^R$}
    \AxiomC{$\B$}
    \UnaryInfC{$b^L, \Delta$}
    \RightLabel{\text{ Cut}}
    \BinaryInfC{$\Gamma, \Delta$}
    \DisplayProof
\end{center}
and need to obtain the conclusion $\Gamma, \Delta$ with a cut free proof. 
The proof again works by double induction: first on the size of the cut formula $b$, then on the total size of $\A$ and $\B$:

\begin{lstlisting}
Lemma inner_cut_elim : forall
  (fuelB: nat) 
  (b: Term) (good_fuelB: fuelB >= termSize b)
  (fuelSize: nat)
  (gamma: AnTerm) (delta: AnTerm)
  (A: OLProof (gamma, R b)) (p1: isCutFree A) 
  (B: OLProof (L b, delta)) (p2: isCutFree B) 
  (good_fuelSize: fuelSize >= (Size A + Size B)), 
  {p: OLProof (gamma, delta) | isCutFree p}.
\end{lstlisting}

We then proceed by case analysis on $\A$ and $\B$, and give a specific transformation in each case. For example,
\begin{center}
\begin{tabular}{c }
    \AxiomC{$\A'$}
    \UnaryInfC{$\Gamma, \alpha^R$}
    \RightLabel{\text { RightOr}}
    \UnaryInfC{$\Gamma, (\alpha \lor \beta)^R$}
    \AxiomC{$\B'$}
    \UnaryInfC{$\alpha^L, \Delta$}
    \AxiomC{$\B''$}
    \UnaryInfC{$\beta^L, \Delta$}
    \RightLabel{\text { LeftOr}}
    \BinaryInfC{$(\alpha \lor \beta)^L, \Delta$}
    \RightLabel{\text{ Cut}}
    \BinaryInfC{$\Gamma, \Delta$}
    \DisplayProof 
    
    \\
    
    $\hookrightarrow$ 

    \\
    
    \AxiomC{$\A'$}
    \UnaryInfC{$\Gamma, \alpha^R$}
    \AxiomC{$\B'$}
    \UnaryInfC{$\alpha^L, \Delta$}
    \RightLabel{\text{ Cut}}
    \BinaryInfC{$\Gamma, \Delta$}
    \DisplayProof
    \\[2ex]
\end{tabular}
\end{center}
Every single recursive use of the induction hypothesis has to provide the required proof of decreasing measures corresponding to the fuel properties. 

The main challenge of the proof comes from the sheer size of the case analysis: the Swap step essentially duplicates every other proof step by allowing them to act on the first or second formula, which implies we have to analyse each of $\A$ and $\B$ on 23 cases each, for a total of more than 500 cases. In practice, we can first do the analysis on $\A$, and for some cases the proof is independent of the structure of $\B$. However, there is no way to undo the case analysis on $\A$ when conversely the cases of pattern matching on $\B$ have a proof independent of the structure of $\A$.

Then, some combinations of cases are impossible. For example, it is not possible for $\A$ to conclude with a \lstinline|RightAnd| and $\B$ with \lstinline|LeftOr1|, as the cut formula $b$ would then need to be both a conjunction and a disjunction. Thanks to using dependent types to define the proof system, those cases are automatically eliminated.
The paper proof can afford a lot of reasoning by symmetry. There are particular combined symmetries between left and right rules, meet and join, swapped and non-swapped cases, $\A$ and $\B$, etc, which can easily be informally treated in a paper proof but not in Coq. In the end, the formal proof contains around 200 cases. 

Among these cases, we caught one in particular that was not properly considered in the paper proof of \cite{guilloudOrthologicAxioms2024}. Indeed, the example reduction above would fail in the presence of an implicit contraction due to set semantics. Written with an explicit contraction, this is:
\begin{center}
\begin{tabular}{c }
    \AxiomC{$\A'$}
    \UnaryInfC{$\Gamma, \alpha^R$}
    \RightLabel{\text { RightOr\hspace{-2em}}}
    \UnaryInfC{$\Gamma, (\alpha \lor \beta)^R$}
    \AxiomC{$\B'$}
    \UnaryInfC{$\alpha^L, (\alpha \lor \beta)^L$}
    \AxiomC{$\B''$}
    \UnaryInfC{$\beta^L, (\alpha \lor \beta)^L$}
    \RightLabel{\text { LeftOr}}
    \BinaryInfC{$(\alpha \lor \beta)^L,(\alpha \lor \beta)^L$}
    \RightLabel{\text { Contract}}
    \UnaryInfC{$(\alpha \lor \beta)^L, N$}
    \RightLabel{\text{ Cut}}
    \BinaryInfC{$\Gamma, N$}
    \DisplayProof
    \\[2ex]
\end{tabular}
\end{center}
And the transformation above would yield $\Gamma, (\alpha \lor \beta)^L$ instead of $\Gamma, N$, and hence is not correct. In this case, a correct transformation is:

\begin{center}
    \AxiomC{$\A'$}
    \UnaryInfC{$\Gamma, \alpha^R$}
    \AxiomC{$\A'$}
    \UnaryInfC{$\Gamma, \alpha^R$}
    \RightLabel{\text { RightOr}}
    \UnaryInfC{$\Gamma, (\alpha \lor \beta)^R$}
    \AxiomC{$\B'$}
    \UnaryInfC{$(\alpha \lor \beta)^L, \alpha^L$}
    \RightLabel{\text{ Cut}}
    \BinaryInfC{$\alpha^L, \Gamma$}
    \RightLabel{\text{ Cut}}
    \BinaryInfC{$\Gamma, \Gamma$}
    \RightLabel{\text{ Contract}}
    \UnaryInfC{$\Gamma, N$}
    \DisplayProof
\end{center}
The topmost cut is justified by induction because its proof is smaller. The second, however, isn't: after recursive elimination from the cut above, we can't guarantee that the new proof is smaller. Instead, this cut needs to be justified by outer induction on the size of the cut formula.

\section{Decision Procedure for Orthologic}
\label{sect:tactic}
The \textit{word problem} for ortholattices consists in deciding, for arbitrary terms over $(\land, \lor, \neg)$ $s$ and $t$, if $s=t$ in all ortholattices. Cases of particular interest involve deciding if $s=1$ (validity) or $s=0$ (unsatisfiability). Since $s=t \iff s \leq t$ \& $t\leq s$, and vice-versa $s \leq t \iff s = s \land t$, we can equivalently see the word problem as deciding inequality between arbitrary words. By soundness and completeness of the orthologic proof system, this is in turn equivalent to deciding if the sequent $s^L, t^R$ has a proof. 

This is decidable with a recursive backward proof search procedure, as in~\cite{guilloudOrthologicAxioms2024}. Given a sequent $\Gamma, \Delta$, try to apply all the rules  which can conclude with $\Gamma, \Delta$, and recursively solve the premise. Since orthologic admits cut elimination, we never need to consider the \texttt{Cut} step. Moreover, because the orthologic proof systems admits the subformula property, we also know that the depth of a proof of a sequent of size $n$ will never exceed $4 n^2$.
We implement this proof search procedure as a Coq function in \autoref{lst:decisionAlgo}, and show soundness and completeness with respect to existence of proof.

\begin{figure*}[!t]
    \begin{lstlisting}[label=lst:decisionAlgo, caption={Decision procedure for the \textit{word problem for ortholattices}.}]
Fixpoint decideOL_base (fuel: nat) (g d: AnTerm) : bool :=
  match fuel with
  | 0 => false
  | S n =>
    match (g, d) with 
    | (L (Var a), R (Var b) )  => (Pos.eqb a b) | _ => false (* Hyp *)
    end || (
    decideOL_base n g N || ( (* Weaken *)
    match d with 
    | N => decideOL_base n g g | _ => false (* Contract *)
    end || (
    match g with 
    | L (Meet a b) => decideOL_base n (L a) d | _ => false (* LeftAnd1 *)
    end || (
    match g with
    | L (Meet a b) => decideOL_base n (L b) d | _ => false (* LeftAnd2 *)
    end || (
    match g with 
    | L (Join a b) => decideOL_base n (L a) d && decideOL_base n (L b) d | _ => false (* LeftJoin *)
    end || (
    match g with 
    | L (Not a) => decideOL_base n (R a) d | _ => false (* LeftNot*)
    end || (
    ... (* Symmetric right cases *)
    || (
    decideOL_base n d g (* Swap *)
    )))))))))))
  end.
    \end{lstlisting}
\end{figure*}

The soundness of this algorithm, which is required for obtaining a reflection tactic, is expressed as
\begin{lstlisting}
Theorem decideOL_base_correct  : 
    forall n g d, 
    (decideOL_ n g d) = true -> 
    exists _: (OLProof (g, d)).
\end{lstlisting}

\subsection*{Reflection}
Suppose we have a goal $s \leq t$ or $s = t$, where $s$ and $t$ are expressions in an arbitrary ortholattice, for example:
\begin{lstlisting}
Lemma example a b: a && negb a = negb b && (a && b).
\end{lstlisting}
We want to solve this problem by executing the above algorithm, which we proved correct. First, we split $s=t$ in two independent inequalities. Then, we \textit{reflect} the problem into the language of orthologic expressions. To do so, we find two terms \lstinline|s'| and \lstinline|t'| as well as a (finite) function \lstinline|f: nat -> bool| such that \lstinline|eval s' f| is convertible to \lstinline|s| and \lstinline|eval t' f| is convertible to \lstinline|t|.
The inequality in the goal is then convertible to:
\begin{lstlisting}
eval ((Var 0) ∩ ¬ (Var 0)) f      <= 
eval (¬ (Var 1) ∩ ((Var 0) ∩ (Var 1))) f.
\end{lstlisting}

We can then apply a lemma derived from the soundness of the decision procedure with respect to the proof system, and of the proof system with respect to the class of ortholattices to reduce the problem to
\begin{lstlisting}
decideOL (L ((Var 0) ∩ ¬ (Var 0)))
         (R (¬ (Var 1) ∩ ((Var 0) ∩ (Var 1))))
    = true.
\end{lstlisting}
Since the left hand-side is convertible to true, this is convertible to \lstinline|true=true|. Effectively, we have offloaded the burden of proof to the evaluator inside Coq's kernel. 

To compute \lstinline|f|, we put all the leaves of $s$ and $t$ in a list \lstinline|env|. A leaf is a subexpression that is neither a meet, a join or a negation. These leaves will correspond to variables in \lstinline|s'| and \lstinline|t'|. $f$ is then the function that maps $n$ to the element in the list at place $n$. The reification tactic then uses this list to compute the terms \lstinline|s'| and \lstinline|t'|.

\iffalse
A somewhat tricky aspect of the reification is its sensibility to aliases. For example, we want our tactic to work both on

\begin{lstlisting}
    a && negb a = negb b && (a && b)
\end{lstlisting}
which uses the language of the specific ortholattice \lstinline|bool|, but also on
\begin{lstlisting}
    a ∩ ¬ a = ¬ b ∩ (a ∩ b)
\end{lstlisting}
which uses the equivalent but different language of arbitrary ortholattices. Here,\lstinline{ ¬ }is itself a notation for \lstinline{@neg OL}, where \lstinline| OL | is the specific ortholattice of \lstinline|a| and \lstinline|b|

The \lstinline|convertible| tactic in \lstinline|reify_term| checks if an expression \lstinline|op| is convertible to \lstinline|@meet OL|, \lstinline|@join OL| or \lstinline|@neg OL|. If so, it recursively computes the reification of the subterms.
The \lstinline|convertible| tactic is implemented so that it does nothing, but fails if its two arguments are not convertible:
\begin{lstlisting}
Ltac convertible x y := 
    constr:(eq_refl x : x = y).
\end{lstlisting}
\fi

We mechanize the entire process as the tactic \lstinline|solve_OL|, which can solve, for example:
\begin{lstlisting}
Lemma example a b: 
    a && negb a = negb b && (a && b).
Proof. solve_OL BoolOL. Qed.
\end{lstlisting}

\section{Verified Memoization}
\label{sect:memoization}

The naively implemented proof search procedure has worst-case runtime complexity of up to $\mathcal{O}(2^{n^2})$, making it unusable in practice. We need to optimize it. 
First, observe that some proof steps, when applied backward, are not merely sufficient conditions but also necessary. For example, consider a sequent of the form
 $\Gamma, (\phi_1 \land \phi_2)^L$.
It is a theorem of lattices that $s \leq t_1 \land t_2 \iff s \leq t_1\ \&\ s \leq t_2$, corresponding to the \texttt{RightAnd} step. Hence, for $\Gamma, (s\land t)^R$ to have a proof, it is not only sufficient but also necessary that both $\Gamma, s^R$ and $\Gamma, t^R$ have a proof.
It follows that if \texttt{RightAnd} is applicable, we do not need to try other steps, and similarly with \texttt{LeftOr}, \texttt{Hyp}, \texttt{RightNot} and \texttt{LeftNot}.
We implement a second version of the algorithm, names \lstinline|decideOL_opti| implementing these optimizations, prove its soundness and implement a corresponding reflection tactic \lstinline|solve_OL_opti|. While these optimizations don't influence the theoretic complexity, they can still reduce runtime in practice.

To obtain a polynomial time procedure, we need to implement \textit{memoization}. Observe that since orthologic admits cut elimination, it also admits the subformula property:

\begin{theorem}[Subformula Property, \cite{guilloudOrthologicAxioms2024}]
If an orthologic sequent $\Gamma, \Delta$  has a proof, then it has a proof where only subformulas of $\Gamma$ and $\Delta$ appear.
\end{theorem}

Let $n$ be the size of the input sequent, i.e. the number of subformulas. The number of subformulas of the initial input sequent is, and there can only exist at most $\mathcal O(n^2)$ different sequents built from these subformulas. Hence, in a full run of \lstinline|decideOL_bool|, there can only be at most $\mathcal O(n^2)$ unique recursive calls. Using memoization, we can ensure that the body of the program is never executed more than $\mathcal{O}(n^2)$ times.

We implement this version of the algorithm using the \textit{state monad} paradigm. Let \lstinline|MemoMap| be some type of maps with keys in \lstinline|(AnTerm * AnTerm)| and values in \lstinline|bool|. The function \lstinline|decideOL_memo| then returns an object of type \lstinline| MemoMap -> (bool, MemoMap)|. Boolean conjunctions and disjunctions are modified to compute the results in series, as in:
\begin{lstlisting}
Definition mor (left : MemoMap -> (bool * MemoMap)) 
               (right : MemoMap -> (bool * MemoMap))  :=
  fun (memo : MemoMap) => 
    match left memo with
    | (true, m) => (true, m)
    | (false, m) => right m
    end.
\end{lstlisting}
and the decision algorithm is modified as follows:
\begin{lstlisting}
Fixpoint decideOL_memo
    (fuel: nat) 
    (g d: AnTerm) 
    (memo: MemoMap) : (bool * MemoMap) :=
  match find (g, d) memo with
  | Some b => (b, memo)
  | None => let (b, m) := ... in
  (b, update_map (g, d) b m)
  end.
\end{lstlisting}

The correctness of the algorithm is expressed as equivalence with the non-memoized version of the algorithm, and relative to the correctness of the map given as input. A map is correct if and only if for every key $k$ with a value of \lstinline|true|, the non-memoized version of the algorithm  returns \lstinline|true| on $k$.
\begin{lstlisting}
Definition memomap_correct (l: MemoMap) :=  forall g d, 
  match find (g, d) l with
  | Some (_, true) => exists n,  (decideOL_base n g d = true)
  | _ => True
  end.
\end{lstlisting}
In particular, the empty map is correct. Note that the two versions of the algorithm are not necessarily equivalent for an arbitrary allowance of fuel: the memoized version might require less. We prove the soundness of the memoized algorithm\footnote{Completeness is not necessary to obtain a reflection-based tactic}:
\begin{lstlisting}
Theorem decideOL_memo_correct  : 
  forall n g d l, 
  (memomap_correct l) -> 
  (memomap_correct (snd (decideOL_fmap n g d l))) /\
  (((fst (decideOL_fmap n g d l)) = true) ->  exists n0, (decideOL_base n0 g d) = true).
\end{lstlisting}

The induction has to be performed jointly on the statement that the returned map is correct, and that the returned truth value is the same as that of the original algorithm. The proof again proceeds with a large case analysis, which required significant custom automation and side lemmas about the monadic structure of the memoization, but was carried completely independently of the orthologic proof system.

We first implement the algorithm using a list of pairs as a base for the memoization map. However, lookup inside a list takes time linear in its size. As explained above, the number of stored values is quadratic in the size of the input. Moreover, checking equality between the input and the elements of the list costs an additional linear factor. Overall, the complexity of lookup is cubic, for a total runtime of $\mathcal{O}(n^5)$. 

We then implement a second version, using AVL maps from the Coq standard library. This requires to define a total order on \lstinline|anTerm|, which we do following the usual total order on labelled ordered trees. Sorted maps such as AVL maps only require a logarithmic number of comparison, down from quadratic with a list-based implementation. However, comparison still takes linear time, for a total time complexity of $\mathcal{O}(n^3\log n)$. 

We obtain hence two more versions of the algorithm, \lstinline|decideOL_memo| and \lstinline|decideOL_fmap|, and two corresponding tactics \lstinline|solve_OL_memo| and \lstinline|solveOL_fmap|.

\section{Reference Equality}
\label{sect:pointers}
Lookup in a map requires deciding either equality or ordering between two terms, which takes linear time in the size of the terms and even in a Hashmap, equality checking is necessary to avoid collisions. 
In the previous two algorithms, either based on lists or on AVL maps, equality is structural: two terms are equal if they have the same constructor and recursively their arguments are equal. However, this is \textit{too strong} for memoization. 

First, observe that if we replace this notion of equality by a strictly weaker notion relation, the algorithm is still sound, and we only risk losing some of the benefits of memoization.

Then recall that by the subformula property, the algorithm only ever sees the $\mathcal{O}(n^2)$ different subnodes of the original input. Assigning a different binary identifier to each of these nodes requires only $\mathcal{O}(\log n)$ bits for each identifier. Then, if two terms have the same identifier, they must be structurally equal. However, the converse does not hold in general: if the original input contains multiple copies of the same subtree, they will be assigned different identifiers. 

This corresponds, in imperative programming, to \textit{pointer equality}. Checking if two objects have the same location in memory is a sound approximation to deciding if they are structurally equal.

Formally, we defined an extended version of the datatype of terms:
\begin{lstlisting}
Inductive TermPointer : Set :=
  | VarP : nat -> Pointer -> TermPointer
  | MeetP : TermPointer -> TermPointer -> Pointer -> TermPointer
  | JoinP : TermPointer -> TermPointer -> Pointer -> TermPointer
  | NotP : TermPointer -> Pointer -> TermPointer.
\end{lstlisting}
where \lstinline|Pointer := positive| is the type of binary positive numbers.
We define the projection onto regular terms \lstinline|ForgetPointer : TermPointer -> Term| and a getter \lstinline|GetPointer : TermPointer -> Pointer| as expected.
We extends pointers to annotated pointers:
\begin{lstlisting}
Inductive AnPointer : Set :=
  | NP : AnPointer
  | LP : Pointer -> AnPointer
  | RP : Pointer -> AnPointer.
\end{lstlisting}
and \lstinline|TermPointer| to \lstinline{AnTermPointer} similarly. For \lstinline|g| an \lstinline|AnTermPointer| (that is, a term with a pointer and a left or right annotation), we note \lstinline|[[g]]| its corresponding \lstinline|AnPointer|. We again use AVL maps, but this time with keys being pairs of \lstinline|AnPointer|. The algorithm is modified accordingly:
\begin{lstlisting}
Fixpoint decideOL_pointers 
    (fuel: positive) 
    (g d: AnTermPointer) 
    (memo: MemoMap) : (bool * MemoMap) :=
  match M.find ([[g]], [[d]]) memo with
  | Some b => (b, memo)
  | None => let (b, m) := ... in
  (b, AnPointerPairAVLMap.add ([[g]], [[d]]) b m)
  end.
\end{lstlisting}

The correctness of the algorithm of course depends on how pointers are assigned. If two different terms are assigned the same pointer, the algorithm will not be correct. Formally, \lstinline|GetPointer| must be injective on the domain of all subterms of the input. In the correctness theorem, it is convenient to express this condition as the existence of a function \lstinline|f: Pointer -> TermPointer|, corresponding to address lookup, which is left and right inverse to \lstinline|GetPointer| on all subterms of the input. 

Proving the correctness of the algorithm extends with moderate effort from the previous algorithm with some additional side lemmas and boilerplate. However, showing that our pointer assignment is correct was much harder. We assign pointers with depth-first, preorder traversal of the syntactic tree.
To construct the required inverse function, we first compute the list of subterms of a term, and map a pointer  to the first term of the list with this pointer. It seems very obvious from the definition of \lstinline|add_pointer| that there exists only one such term and hence that the two functions are inverse of each other, yet this was surprisingly difficult to prove, and required a lot of intermediate lemmas about the structure of subterms, the monotonicity of \lstinline|add_pointer| along subterms, correspondence between the pointer and non-pointer version of terms, and more.

It was however entirely independent of the orthologic proof system, and only depended upon the number and arity of constructors of orthologic terms. Hence, the theorems trivially transfer to any other algebraic datatype.

\section{Proof-Producing Tactics and Validity Solvers}
\label{sect:ocaml}

An alternative to verified decision procedures is \textit{proof-producing decision procedures}, which also allows to implement tactics for proof assistant. Instead of relying on a soundness theorem, a proof-producing tactic will compute a proof (in Coq, a proof term) that is then checked by the logical kernel.

We implemented a proof-producing version of the proof search algorithm of \cite{guilloudOrthologicAxioms2024} in OCaml, and make it available as a Coq tactic, which we refer to as \lstinline|olcert_goal|. 
In theory, this should be less efficient, as producing a proof takes additional time, but in practice an OCaml implementation leveraging in particular mutability is a faster computation method than reducing lambda-term. Compared to \lstinline|solveOL_pointer| it has the practical benefits of persistence of its memoization across calls, as the pointers are assigned once and do not change and the memoization is global. Moreover, when possible, we memoize results as fields of the term nodes rather than in a map, as retrieval is more efficient. 

On the other hand, the proof term produced has size up to $\mathcal O(n^2)$, while the proof term of \lstinline|solveOL_pointer| has constant size.
Moreover, the proof-producing tactic is not verified, so it is not guaranteed that it does not contain a bug and outputs an incorrect proof. In fact, because Coq's logic is rather complex and its inner workings hard to work with (and sparsely documented), it is quite likely that our implementation contains bugs on edge cases, for example involving universes or unification variables.

We then develop a tactic \lstinline|olnormalize|, described in \cite{guilloudFormulaNormalizationsVerification2023}. An OCaml algorithm computes the normal form $f'$ of a formula $f$, and we then use either \lstinline|solveOL_pointer| or \lstinline|olcert_goal| to prove $f = f'$. We believe this tactic has the most impact for automation, as \cite{guilloudFormulaNormalizationsVerification2023} demonstrated it can significantly reduce the size of formulas in practical contexts. For example on the first benchmark of next section:
\begin{lstlisting}
Theorem test_tauto02_0 (x0 x1: bool) :
  ! (((! x0 && ! x0) || (x0 && x1) || (x0 && ! x1) || (! x0 && ! x0)) && 
    ((x1 && x0) || (! x0 && x1) || (! x0 && ! x1) || (! x0 && x1))) 
    = 
  true
. Proof.
  olnormalize. 
  (* (! x1 || ! x0) && (x0 || x1) && (x0 || ! x1)   = true*)
\end{lstlisting}

Finally, we use the \lstinline|olnormalize| tactic as basis for a complete tactic for Boolean equality (i.e. validity). This tactic is similar to the DPLL algorithm: it recursively branches on a literal and simplify the result. The simplification is performed using \lstinline|olnormalize|. 
We implement two versions of this tactic, one named \lstinline|oltauto| which uses the reflexive orthologic tactic and one named \lstinline|oltauto_cert|, which uses the proof-producing orthologic tactic.

\section{Evaluation of Resulting Tactics}
\label{sect:bench}
\paragraph{Evaluation of the OL proof search tactics.}
The implementations of the previous section yield six algorithms and corresponding proof tactics for deciding equality in orthologic, each with different time complexity\footnote{Exact complexity of algorithms up to logarithmic factor depends on the precise model of computation; for simplicity, the algorithmic complexity below assume the usual Word RAM model, where checking equality of words takes constant time, even though this does not precisely match Coq's evaluation of terms. }:
\newcommand{\smalltt}[1]{\texttt{\footnotesize{#1}}}
\begin{itemize}
    \item \lstinline|solve_OL| (\smalltt{"OL"}, $\mathcal O (2^n)$)
    \item \lstinline|solve_OL_opti| (\smalltt{"OL+o"}, $\mathcal O (2^n)$)
    \item \lstinline|solve_OL_memo| (\smalltt{"OL+o+l"}, $\widetilde{\mathcal O} (n^5)$)
    \item \lstinline|solve_OL_fmap| (\smalltt{"OL+o+m"}, $\widetilde{\mathcal O} (n^3)$)
    \item \lstinline|solve_OL_pointers| (\smalltt{"OL+o+m+φ"}, $\widetilde{\mathcal O} (n^2)$)
    \item \lstinline|olcert_goal| (\smalltt{"OCaml"}, $\mathcal O (n^2)$)
\end{itemize}
The difference of complexity is not only theoretical, but also highly observable in practice. Consider the family of propositions
$$
\phantom{{}={}}(x_1 || x_2) \&\& (x_3 || x_4) \&\& ... \&\& (x_{n-1} || x_n) 
$$
$$
{}={}(x_2 || x_1) \&\& (x_4 || x_3) \&\& ... \&\& (x_n || x_{n-1})
$$
These equalities are solvable by the laws of ortholattices.  Table~\ref{tab:runtime} shows the time taken by each implementation for $n=30$.  For comparison, we also include the runtime of \lstinline|btauto|, the solver for boolean equalities included in Coq's standard distribution.

\begin{table}
\begin{center}
\begin{tabular}{l@{\hspace{1em}}r@{\hspace{1em}}r}
\hline
 solver & mean & std \\
\hline
btauto & 20.94 & 0.31 \\
OL+o+l+vm & 9.44 & 0.04 \\
OL+o+m+vm & 0.23 & 0.00 \\
OL+o+m+φ+vm & 0.17 & 0.00 \\
OCaml & 0.05 & 0.00 \\
\hline
\end{tabular}
\end{center}
\caption{\label{tab:runtime}Wall clock time required to prove an equality involving 30 variables.}
\end{table}

\begin{figure}[!t]
    \includegraphics[width=\linewidth]{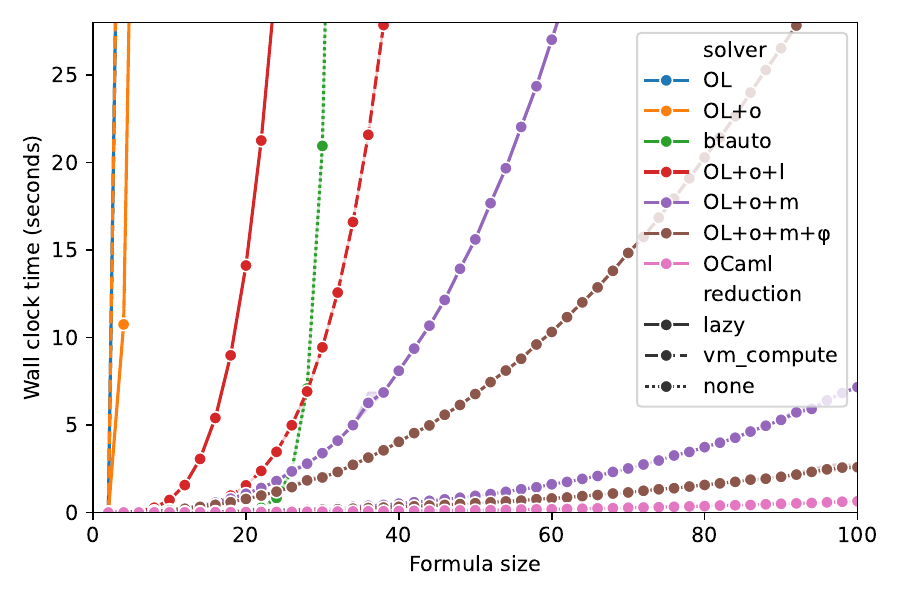}
    \caption{\label{fig:runtime}Wall clock time required to prove a family of equalities with sizes ranging from 2 to 100 variables. Shaded regions indicate 95\% confidence intervals.  Colours indicate which implementation was used; line styles indicate which reduction strategy was used in the final step of the proof by reflection. Benchmarks were run on an Intel Core i9-13900K CPU with 64GB RAM.}
\end{figure}

To confirm the scaling characteristics of our implementation variants, we solved the family of equalities above for sizes ranging from 2 to 100 variables.  Figure~\ref{fig:runtime} shows our results.
We confirm the theoretical complexity by verifying for each tactics  that a polynomial of the right degree fits the measurements.

The proof-producing implementation in OCaml  outperforms the implementation as a Coq function (i.e. in Gallina), but it produces longer proof term, and since its soundness is not proven it may contain bugs. But it also demonstrates the inadequacy of algorithms implemented directly in Coq for practical purposes, as reduction of lambda-terms is not an efficient computational strategy. This suggests that in an ideal world, it should be possible to verify functions implemented in a general-purpose programming language and then trust their output, so that they can run as fast as possible. This was done in some flavour with the Candle proof assistant (a verified implementation of HOL Light) in \cite{abrahamssonFastVerifiedComputation2023}.

\begin{figure}[!t]
    \hspace{-3em}\includegraphics[width=0.4\linewidth]{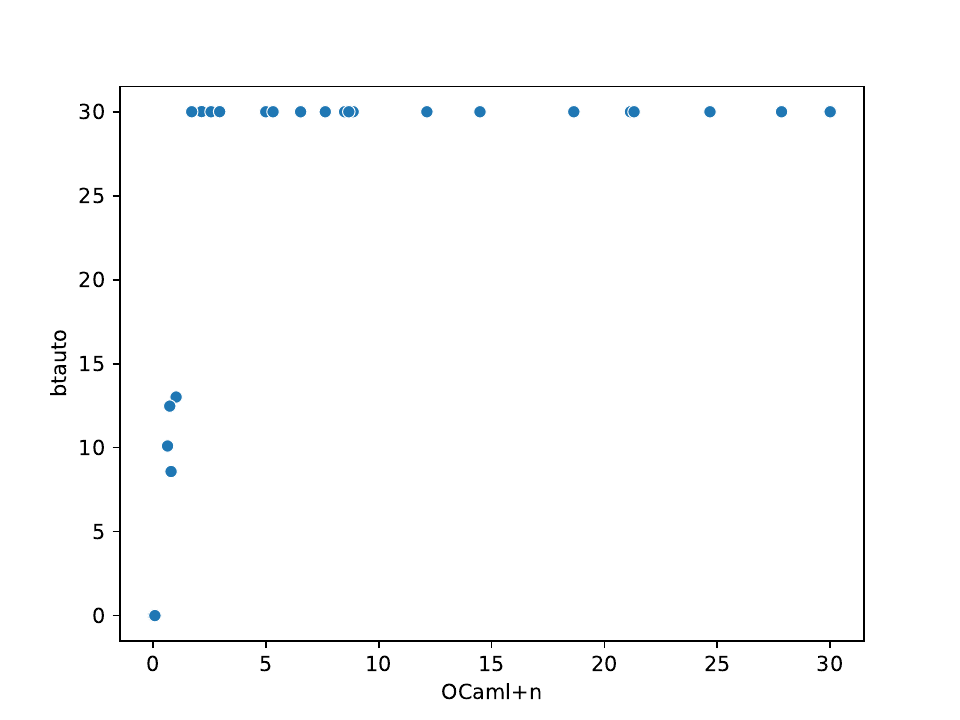}\hspace{-5em}\hfill
    \hspace{-5em}\includegraphics[width=0.4\linewidth]{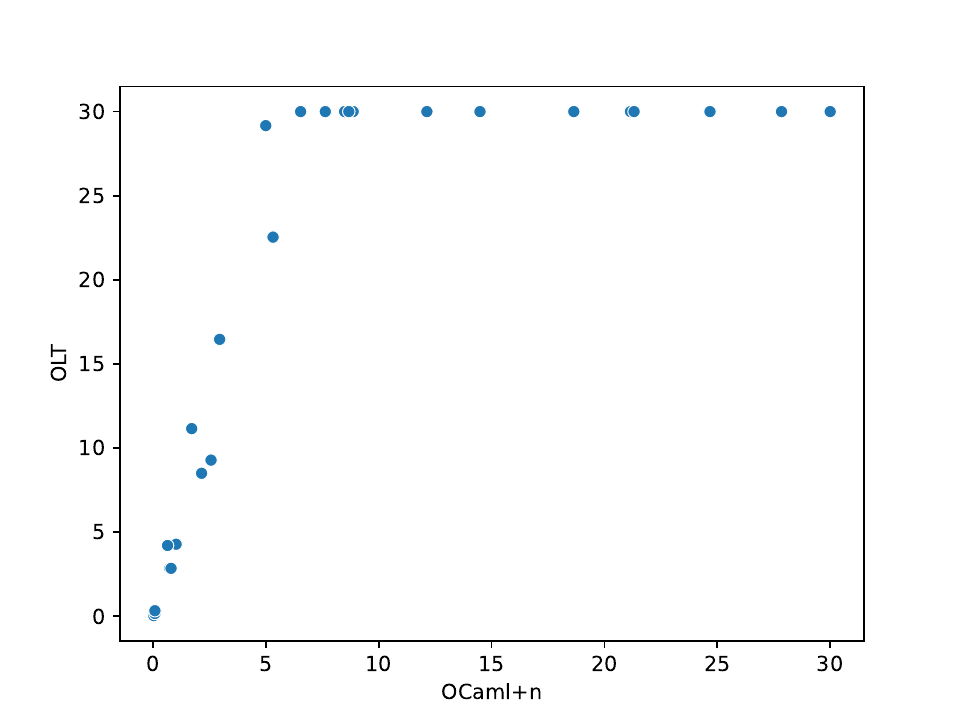}\hspace{-5em}\hfill
    \hspace{-5em}\includegraphics[width=0.4\linewidth]{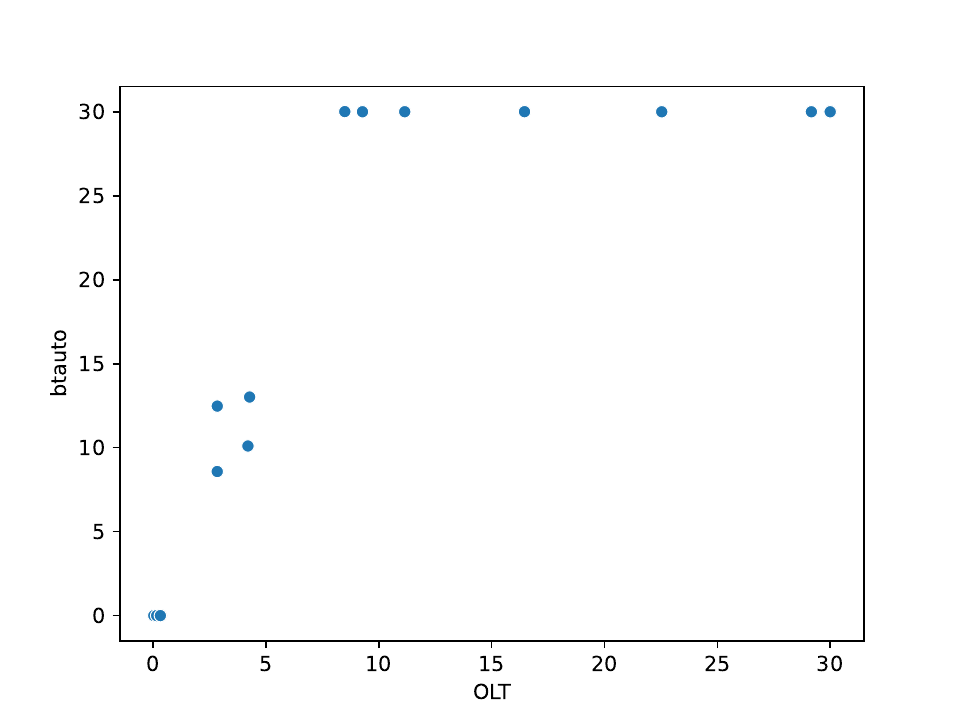}\hspace{-3em}
    \caption{\label{fig:runtime_tauto}Wall clock time required to prove random hard non-clausal formulas. Each point is an experiment and indicates the running time for two of the three solvers. }
\end{figure}

\paragraph{Evaluating propositional solvers.}
To judge the tactics \lstinline|oltauto| and \lstinline|oltauto_cert| compared to the standard \lstinline|btauto| solver, we generate a series of formulas according to the techniques of \cite{navarroGenerationHardNonclausal2005}. These formulas have a fixed structure and are believed to be typically hard for SAT solvers. Some of the formulas we generated are valid and some are not, but this does not affect the runtime meaningfully as all three algorithms do not stop early if a dissatisfying assignment is found and still explore the entire search space. We generate a total of 80 random formulas, with up to 20 different variables and 192 literals, and a timeout of 30 seconds. The results appear in \autoref{fig:runtime_tauto} and show that \lstinline|oltauto_cert| (OCaml+n) consistently outperforms \lstinline|oltauto| (OLT), which in turns consistently outperforms \lstinline|btauto|.

All the tactics as well as the code used to generate and run benchmark is available at [redacted].

\section{Conclusion}
We formalized Orthologic in the Coq proof assistant, including soundness and completeness of orthologic sequent calculus with respect to the class of all ortholattices and the cut elimination theorem for orthologic. Doing so, we discovered a missing edge case in the proof from \cite{guilloudOrthologicAxioms2024}.

We implemented and verified a proof search procedure for orthologic, and used reflection to obtain a proof tactic that decides equalities and inequalities in ortholattices. We improved the algorithm with important but typically imperative programming features: memoization and pointer equality, improving the algorithm from exponential to quadratic. We implemented additional specialized tactics enabling orthologic-based reasoning for Coq users, all of which are available in a Coq plugin.

Orthologic is used as a core component of reasoning tools, and the formalization of its key properties increases the trust we can have in existing and future such systems.
Our work also opens for further mechanization of orthologic-based reasoning, such as orthologic with axioms and effectively propositional orthologic \cite{guilloudOrthologicAxioms2024}, interpolation in orthologic \cite{miyazakiPropertiesOrthologics2005,guilloudInterpolationQuantifiersOrtholattices2024}, and more \cite{brunsFreeOrtholattices1976,herrmannVarietiesModularOrtholattices,sherifDecisionProblemOrthomodular1997}.

Finally, verification of memoization and reference equality-based optimizations, which represent a significant part of the formalization effort, are in no specific to orthologic proof search.
They are also of great practical relevance, and necessary to obtain optimal complexity of a wide range of algorithms. We  expect that our formalization can be a useful source to anyone verifying efficient algorithms relying on memoization or reference equality in the future.

\bibliographystyle{plainurl}
\bibliography{sguilloud}

\end{document}